\documentclass[aps,prl,reprint,groupedaddress,amsmath,amssymb, showpacs]{revtex4-1}

\usepackage{graphicx}% Include figure files
\usepackage{dcolumn}% Align table columns on decimal point
\usepackage{bm}% bold math
\usepackage{color}
\usepackage{soul}

\setstcolor{red}

\newcommand{\ums}{~$\mu$m~}
\newcommand{\um}{~$\mu$m}

\begin{document}
\title{Long-Range Plasmon Assisted Energy Transfer \\ Between Fluorescent Emitters}

\author{D. Bouchet}
\affiliation{ESPCI ParisTech, PSL Research University, CNRS, Institut Langevin, 1 rue Jussieu, F-75005, Paris, France}
\author{D. Cao}
\affiliation{ESPCI ParisTech, PSL Research University, CNRS, Institut Langevin, 1 rue Jussieu, F-75005, Paris, France}
\author{R. Carminati}
\affiliation{ESPCI ParisTech, PSL Research University, CNRS, Institut Langevin, 1 rue Jussieu, F-75005, Paris, France}
\author{Y. De Wilde}
\affiliation{ESPCI ParisTech, PSL Research University, CNRS, Institut Langevin, 1 rue Jussieu, F-75005, Paris, France}
\author{V. Krachmalnicoff}
\email{valentina.krachmalnicoff@espci.fr}
\affiliation{ESPCI ParisTech, PSL Research University, CNRS, Institut Langevin, 1 rue Jussieu, F-75005, Paris, France}

\date{\today}

\begin{abstract}
We demonstrate plasmon assisted energy transfer between fluorophores located at distances up to $7$\ums on the top of a thin silver film. Thanks to the strong confinement and large propagation length of surface plasmon polaritons, the range of the energy transfer is almost two orders of magnitude larger than the values reported in the literature so far. The parameters driving the energy transfer range are thoroughly characterized and are in very good agreement with theoretically expected values. %This work shows the benefit of plasmonic structures for long-range energy transfer. 
\end{abstract}

\maketitle

Coupling of fluorescent emitters has been a topic of major interest in research over the past decades. The relevance of this subject stems from its many possible exploitations. In the domains of chemical-physics and life science, short-range dipole-dipole interaction (F\"orster Resonant Energy Transfer, FRET) between fluorophores has been widely employed to study phenomena occurring over distances of the order of $10$~nm \cite{Forster1959, MedintzBook, Ha2001}.

Coupling between fluorophores can be tailored by shaping their electromagnetic environment. Together with the emergence of nanophotonics, several studies about the influence of a nanostructured environment on fluorescent emission have come to light \cite{Carminati2015}. In particular, it has been demonstrated that metallic structures can lead to the control of the rate, the intensity and the directionality of fluorescent emission \cite{Koenderink2015, Curto2010}, and single photon generation of single plasmons has been observed and studied \cite{Tame2013, Wrachtrup2009, Fakonas2014}. Plasmonic structures are attractive because, due to the high confinement of plasmon modes, they can ensure a very efficient coupling between the emitter and the environment, as demonstrated by the observation of high Purcell factors \cite{Akselrod2014,Belacel2013}. Furthermore, plasmons can propagate over distances going from a few microns to several millimeters, depending on the material and the wavelength. 

While the interaction between plasmons and one fluorescent emitter has already been widely studied, the influence of a plasmonic environment on the coupling between different fluorophores has been explored only in very specific situations involving FRET \cite{Leitner1988, Dood2005, Nakamura2006, Zhang2007, Komarala2008, Reil2008, Faessler2011, Blum2012, Ghenuche2014, Torres2015}.
In this regime, dipole-dipole interaction restricts energy transfer to distances of a few nanometers. In this Letter, we study the interaction between fluorophores in a plasmonic environment from a different perspective. We take benefit of the confined propagation of surface plasmons polaritons (SPP) along the interface between a metal and a dielectric to realize the transfer of energy between fluorophores on distances which are much larger than what would be allowed by FRET. 
We report on the observation of plasmon assisted energy transfer between ensembles of fluorescent molecules lying on top of a continuous silver film over distances up to $7$\um. The first ensemble (donor) is excited by a laser and then decays by launching a SPP propagating on the top surface of the metallic layer. The energy carried by the plasmon is then absorbed by the second ensemble of fluorophores (acceptor) embedded in a thin polymer layer deposited on top of the metal. Fluorescence photons emitted by the acceptor are the signature of the plasmon assisted energy transfer between the donor and the acceptor. On the basis of a simple theoretical model we estimated the energy transfer efficiency to be enhanced by the presence of the plasmonic film by up to a factor of 30.

\begin{figure}[ht]
\includegraphics[width=8cm]{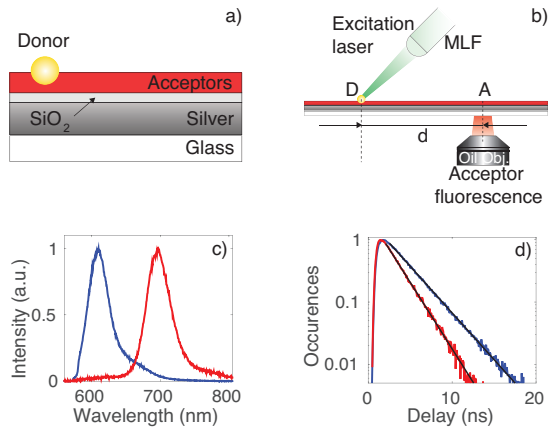}
\caption{\label{fig:Sample} a) Sketch of the sample. b) Scheme of the experimental setup. MLF: micro-lensed fiber, D: donor, A: acceptor. c) Fluorescence spectrum of the donor only (blue) and of the acceptor only (red) on the top of a $50$~nm thick silver film with a $10$~nm silica spacer. d) Fluorescence decay rate of a donor bead on the silver film in the presence (red) and in the absence of the acceptors (blue).}
\end{figure}

%\paragraph{Sample and experimental setup}
The sample, shown in Fig.~\ref{fig:Sample}a, is made of a $50$~nm thick silver layer deposited on a glass coverslip and then covered by a layer of SiO$_2$ (thickness $10$~nm) which protects the metal against oxidation and  prevents quenching of the fluorophores by the metal. An aqueous solution containing polyvinyl alcohol (PVA), acceptor molecules (Atto665, AttoTec Gmbh) and donor polystyrene fluorescent beads (Red FluoSpheres, $100$~nm diameter, Invitrogen) was then spin-coated on  the silica layer. The parameters of the spin-coating process and the concentration of acceptors and donors in the initial solution are chosen in order to have a few isolated donor beads, surrounded by a homogeneous layer of acceptors, about $50$~nm thick. The distance between two close-by donors is such that a single donor bead can be addressed easily with the excitation laser.        
Care has been taken that the donor emission spectrum and the acceptor absorption spectrum have a good overlap, which is needed for the energy transfer to occur. Moreover, since the aim is to realize the transfer via a SPP, the spectral overlap of the donor-acceptor pair has been chosen such that it occurs in a region of the spectrum for which the plasmon can propagate over distances up to several microns.

In the experiment, we excite the donor bead with a supercontinuum pulsed laser (Fianium SC450,  repetition rate $10$~MHz), filtered at $532$~nm (filter bandwidth $20$~nm). The excitation laser, as shown in Fig. \ref{fig:Sample}, is focused on the bead over a spot of $1.5$\ums by means of a micro-lensed fiber (Nanonics Imaging Ltd., working distance $4$\um), mounted on a three axis piezoelectric nanopositioner. Efficient collection of acceptor fluorescence is ensured   
by an oil immersion microscope objective ($\times100$, N.A.=1.4), located below the sample (see Fig. \ref{fig:Sample}b). Fluorescence photons are then filtered by a dichroic mirror and detected either by a fibered spectrometer (Acton SP2300, Princeton Instruments, fiber core diameter $50$\um) or, after passing through a confocal pinhole (diameter $50$\um), by an avalanche photodiode (MPD PDM-series) in order to perform decay rate measurements (acquisition board PicoQuant HydraHarp400). In either case, the signal which is measured arises from a 500 nm sized region of the sample which is optically conjugated with the detection system. The use of the microlensed fiber allows us to separate the excitation and the detection regions by a very well controlled distance $d$ going up to several microns. Indeed, the position of the microscope objective is fixed while both the sample and the microlensed fiber can be translated. To measure the acceptor fluorescence as a function of $d$, we move the donor and the microlensed fiber together by a distance $d$ with respect to the microscope objective. Fluorescence of the donor collected through the microlensed fiber is monitored on an avalanche photodiode and maximized for each distance. An EM-CCD camera allow us to record wide-field images of the sample.   
Fluorescence emission spectra of acceptors only on silver and of a donor only on silver are reported in Fig. \ref{fig:Sample}c in red and blue respectively. Moreover, in order to check the influence on the donor of the presence of acceptors in its proximity, we compared the fluorescence decay of the donor on silver in the presence and in the absence of the acceptors, when the excitation spot coincides with the fluorescence collection region (i.e. $d=0$\ums in Fig. \ref{fig:Sample}b). This measurement is reported in Fig. \ref{fig:Sample}d and shows that the decay rate of the donor in the presence of the acceptors (red data) is larger than without acceptors (blue data) by a factor of $1.4$ (see \cite{Supplemental_Material} for complementary information). This effect is due to the dipole-dipole interaction between the donor and the acceptors surrounding the bead, which is responsible for a short-range energy transfer (FRET) occurring between molecules located at distances of the order of ten nanometers. 
 
\begin{figure}[b]
\includegraphics[width=8cm]{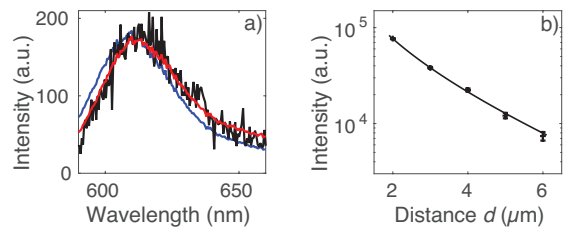}
\caption{\label{fig:donor} a) Normalized spectra of the radiative losses of the plasmon launched by the donor due to scattering by surface roughness. Blue and black curves: experimental data for $d=0$ and $d=5$\ums respectively. Red curve: fit to the experimental data at $d=5$\ums taking into account the dependence, on the wavelength, of the dielectric constant of silver. b) Scattered intensity as a function of the distance $d$. Black points: experimental data, black solid line: fit to the experimental data (see text for details). Error bars have been calculated by taking into account the statistical error on the entries in each bin of the spectrum histogram.}
\end{figure}

%\paragraph{Experimental results - donor only}
In order to characterize the propagation of the plasmon launched by the donor on the silver film, we monitored the radiative losses of the plasmon for several distances $d$ up to $6$\um. The sample used is made as the one previously described except that the acceptor molecules are absent in the PVA layer. Experimental results are shown in Fig. \ref{fig:donor}. 
The wavelength dependence of the dielectric constant of silver induces a dependence of the propagation length of the plasmon on the wavelength. Since longer wavelengths of the spectrum propagate over longer distances, the plasmon spectrum is red-shifted during the propagation. This appears clearly in Fig. \ref{fig:donor}a that shows, in blue, the spectrum of the donor at $d=0$\ums and, in black, the spectrum of plasmonic radiative losses detected at $d=5$\um. In order to confirm the origin of the observed red-shift, we fit the experimental spectra, for several distances $d$, with a function given by the product between the spectrum of the donor on silver at $d=0$\ums and the decreasing exponential decay behavior of the plasmon propagating along the interface between a semi-infinite silver layer and a semi-infinite PVA layer ($\epsilon=2.25$). The dependence of the dielectric constant of the silver film on the wavelength is taken into account \cite{Rakic1998}, on the basis of the Drude model with five interband transitions. The result of the fit for $d=5$\ums is the red curve plotted in Fig. \ref{fig:donor}a, which is in very good agreement with the experimental data. We have verified that a good agreement is found when fitting the data for distances up to $6$\um. At larger distances the signal to noise ratio in the spectroscopic measurements is of the order of unity which prevents one to perform any reliable analysis of the data. The spectrally averaged propagation distance of the SPPs excited by the donor, $l_{D}$, can be estimated from the experimental spectra recorded as a function of the distance $d$. Fig. \ref{fig:donor}b shows a plot of the total intensity of plasmonic radiative losses as a function of $d$ together with a fit (solid line in Fig. \ref{fig:donor}b) using the function $A/d \, \exp (-d/l_{D})$, where $A$ is a normalization constant. The fit results in a propagation length of $3.4 \pm 0.1$\um. This is slightly smaller than the theoretically expected value $l_{D}^{th} = 3.8$\ums estimated by calculating the average of the SPP propagation length over all the donor fluorescence spectrum range, weighted by the fluorescence intensity at each wavelength. The slight difference is likely due to the theoretical estimation in which losses due to surface roughness are not accounted for. 

%\paragraph{Experimental results - donor and acceptors}
After characterizing the propagation of the plasmon launched by the donor, we studied the occurrence of the energy transfer between the plasmon and the acceptors embedded in the PVA film. For that purpose, we measured the spectrum at different distances $d$ from the donor position on the sample sketched in Fig. \ref{fig:Sample}a. To have a thorough understanding of the measurements, the various processes that could contribute to the signal collected at distance $d$ from the donor must all be appropriately considered.   
\begin{figure}[b]
\includegraphics[width=7cm]{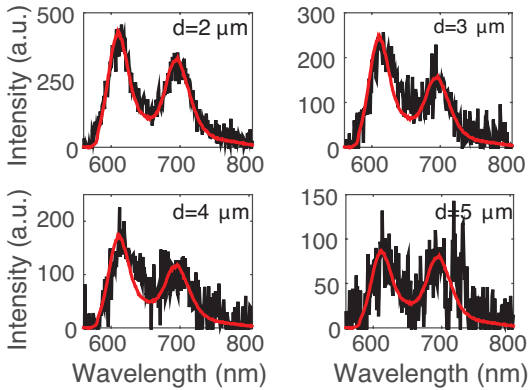}
\caption{\label{fig:spectrum_data} Black points: Spectra measured at a distance $d$ from the donor ranging from $2$\ums to $5$\um, with an integration time of $200$~s. The red curve is the result of a fit of the experimental data with a function which is the sum of the acceptor fluorescence spectrum and the plasmon spectrum.}
\end{figure}
First, due to the finite size of the excitation laser spot (diameter of $1.5$\um) acceptors located in the proximity of the donor can be directly excited by the laser. This effect has been minimized by blue detuning the excitation laser with respect to the acceptor absorption spectrum, but it is still present and acts as a background that has to be subtracted from our measurements. Second, acceptors that are directly excited by the laser, may also excite a SPP. As in the case of the plasmon launched by the donor, radiative losses by the plasmon launched by the acceptors can be detected at a distance $d$. Their spectrum will overlap with the acceptor fluorescence peak, by artificially increasing the signal to be measured. Third, since acceptors are present in the vicinity of the donor, short range energy transfer (FRET) occurs. Acceptors excited by this mechanism can decay by launching a SPP with a spectrum overlapping with the acceptor fluorescence spectrum, that can be detected at a distance $d$ due to radiative losses. Since the contribution of the FRET signal is negligible with respect to the signal emitted by the bead, we neglected this effect. In order to account for the contribution of the first and the second processes, we measured, for each distance $d$, the signal produced by the acceptors due to direct laser excitation, in the absence of the donor. This is done by shifting the sample such that only acceptors are excited by the laser. This background signal is subtracted from the signal measured in the presence of the donor. 

The result is shown in Fig. \ref{fig:spectrum_data} (black data) for several distances $d$. The peak centered at $610$~nm is due to radiative losses of the SPP launched by the donor, while the peak centered at $695$~nm is the fluorescence of the acceptors excited by the plasmon. The presence of the second peak is the unambiguous signature of the occurrence of energy transfer through the SPP excitation. For comparison, a control sample made of donors and acceptors on a bare glass substrate shows only short range energy transfer (FRET). The curve shown in red is a fit to the experimental data by means of a function that is the sum of the spectrum of the plasmon excited by the donor as measured on the sample without acceptors (see Fig. \ref{fig:donor}a) and of the spectrum of acceptor fluorescence on silver (shown in Fig. \ref{fig:Sample}c). The fit function has only two free parameters that are normalization constants. The good agreement between the fit result and the experimental data reinforces the conclusions.     
\begin{figure}[ht!]
\includegraphics[width=7cm]{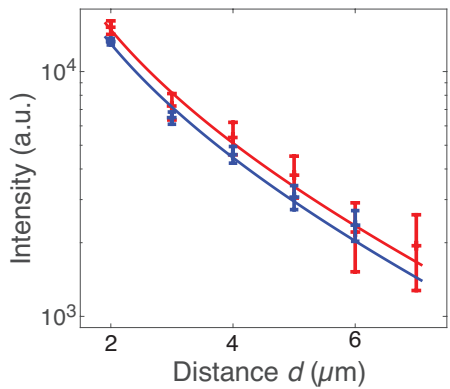}
\caption{\label{fig:energy_transfer} Red: Intensity of the acceptor fluorescence as a function of the distance $d$ from the donor (dots) and expected theoretical dependence (solid line). Blue: Intensity, as a function of the distance $d$, of the plasmon launched by the donor integrated over the region of the spectrum that overlaps with the acceptor absorption, i.e. $635 ~\mathrm{nm} \leq \lambda \leq 680$~nm, (dots) and expected theoretical dependence (solid line). Error bars have been calculated by taking into account the statistical error on the entries in each bin of the spectrum histogram.}
\end{figure}
A second proof of the observation of plasmon mediated energy transfer over distances of up to $7$\ums  is given by the graph shown in red in Fig. \ref{fig:energy_transfer}, where we report the integral of the energy transfer spectra, after subtraction of the plasmon losses signal, as a function of the distance from the donor. As done in Fig. \ref{fig:donor}b, data are fitted with a decreasing exponential function to take into account the exponential decay of the plasmon propagation as a function of the distance. This allows us to measure the energy transfer range $l_{et}=5.4 \pm 0.9$\um. We checked that, for different donors on the same sample and for different acceptor concentrations, the energy transfer range is consistent with the data shown in the manuscript.

At first sight, it might be surprising to measure $l_{et} > l_{D}$. However this can be understood by taking into account the wavelength dependence of the plasmon energy absorbed by the acceptors. Indeed, if we measure $l_D$ by taking into account only the spectral region around the maximum of the acceptor absorption spectrum (i.e. $635~\mathrm{nm} \leq \lambda \leq 680$~nm), we find a propagation length of $5.3 \pm 1$\um. This result agrees with the value of $l_{et}$. Data are plotted in blue in Fig. \ref{fig:energy_transfer} together with the fit result (blue solid lines). This is a further evidence of the fact that acceptors are excited via the SPP launched by the donor.   

The energy transfer rate is defined as the enhancement of the donor decay rate induced by the presence of the acceptors. In the present realization of the experiment, a reliable measurement of the energy transfer rate at long distances is not possible, due to the occurrence of FRET between the donors and the surrounding acceptors. Therefore, we theoretically estimated the plasmon assisted energy transfer efficiency by calculating the Green function at the interface between two semi-infinite media (PVA and silver). As it is reported in the supplemental material, the presence of the metallic film induces an enhancement of up to a factor of 30 with respect to the energy transfer occurring in a homogeneous medium ($n=1.5$) \cite{Supplemental_Material}.
%\paragraph{Conclusion} 
In conclusion, we have demonstrated the occurrence of SPP assisted energy transfer between fluorescent emitters located at distances of up to $7$\ums on the top of a thin silver film. This observation was supported by the measurement of the propagation distance of the SPP on silver, measured on an independent sample and by a very good agreement with the theoretically expected value.

The transfer range which we demonstrate is 60 times larger than the only value reported in the literature so far on plasmon mediated energy transfer between fluorescent emitters \cite{Andrew2004}. In \cite{Andrew2004}, the authors studied the energy transfer between two layers of molecules located on both sides of a silver film, a configuration that differs from the one studied in this Letter. The maximum thickness of the film, fixing the maximum distance of the energy transfer, was 120 nm. 

This work is the first step towards the observation of a plasmon assisted coupling between fluorophores, and opens rich perspectives. Indeed, a more efficient, long-range energy transfer can be reached by shaping the plasmonic support. Energy transfer efficiency can be enhanced by several orders of magnitude by using structured plasmonic supports, such as hybrid plasmon-dielectric waveguides \cite{Marocico2011} and could allow the demonstration of efficient long-range energy transfer between single photon emitters. The realization of optimized substrates for the observation of the energy transfer would open rich perspectives in several domains that are concerned by inter-molecular coupling, spanning from biology to physics. In particular, the use of recently developed techniques for the accurate nanopositioning of fluorescent emitters in the near-field of nanostructured samples \cite{Krachmalnicoff2013, Cao2015} will be required for studies of optimized structures towards the realization of devices.

%The authors thank E. Fort for stimulating discussions, S. Gresillon and O. Loison for helping in sample preparation and A. Souilah for technical support. This work was supported by LABEX WIFI (Laboratory of Excellence ANR-10-LABX-24) within the French Program “Investments for the Future” under reference ANR-10- IDEX-0001-02 PSL* and by the Region Ile-de-France in the framework of DIM Nano-K. 

\begin{acknowledgments}
The authors thank E. Fort and A. Caz{\'e} for stimulating discussions, S. Gresillon and O. Loison for helping in sample preparation and A. Souilah for technical support. This work was supported by LABEX WIFI (Laboratory of Excellence ANR-10-LABX-24) within the French Program “Investments for the Future” under reference ANR-10- IDEX-0001-02 PSL*, by the
Region Ile-de-France in the framework of DIM Nano-K and by the Programme Emergences 2015 of the City of Paris. 
\end{acknowledgments}



\begin{thebibliography}{35}%
\makeatletter
\providecommand \@ifxundefined [1]{%
 \@ifx{#1\undefined}
}%
\providecommand \@ifnum [1]{%
 \ifnum #1\expandafter \@firstoftwo
 \else \expandafter \@secondoftwo
 \fi
}%
\providecommand \@ifx [1]{%
 \ifx #1\expandafter \@firstoftwo
 \else \expandafter \@secondoftwo
 \fi
}%
\providecommand \natexlab [1]{#1}%
\providecommand \enquote  [1]{``#1''}%
\providecommand \bibnamefont  [1]{#1}%
\providecommand \bibfnamefont [1]{#1}%
\providecommand \citenamefont [1]{#1}%
\providecommand \href@noop [0]{\@secondoftwo}%
\providecommand \href [0]{\begingroup \@sanitize@url \@href}%
\providecommand \@href[1]{\@@startlink{#1}\@@href}%
\providecommand \@@href[1]{\endgroup#1\@@endlink}%
\providecommand \@sanitize@url [0]{\catcode `\\12\catcode `\$12\catcode
  `\&12\catcode `\#12\catcode `\^12\catcode `\_12\catcode `\%12\relax}%
\providecommand \@@startlink[1]{}%
\providecommand \@@endlink[0]{}%
\providecommand \url  [0]{\begingroup\@sanitize@url \@url }%
\providecommand \@url [1]{\endgroup\@href {#1}{\urlprefix }}%
\providecommand \urlprefix  [0]{URL }%
\providecommand \Eprint [0]{\href }%
\providecommand \doibase [0]{http://dx.doi.org/}%
\providecommand \selectlanguage [0]{\@gobble}%
\providecommand \bibinfo  [0]{\@secondoftwo}%
\providecommand \bibfield  [0]{\@secondoftwo}%
\providecommand \translation [1]{[#1]}%
\providecommand \BibitemOpen [0]{}%
\providecommand \bibitemStop [0]{}%
\providecommand \bibitemNoStop [0]{.\EOS\space}%
\providecommand \EOS [0]{\spacefactor3000\relax}%
\providecommand \BibitemShut  [1]{\csname bibitem#1\endcsname}%
\let\auto@bib@innerbib\@empty
%</preamble>
\bibitem [{\citenamefont {F{\"o}rster}(1959)}]{Forster1959}%
  \BibitemOpen
  \bibfield  {author} {\bibinfo {author} {\bibfnamefont {T.}~\bibnamefont
  {F{\"o}rster}},\ }\href@noop {} {\bibfield  {journal} {\bibinfo  {journal}
  {Discussions of the Faraday Society}\ }\textbf {\bibinfo {volume} {27}},\
  \bibinfo {pages} {7} (\bibinfo {year} {1959})}\BibitemShut {NoStop}%
\bibitem [{\citenamefont {Medintz}\ and\ \citenamefont
  {Hildebrandt}(2003)}]{MedintzBook}%
  \BibitemOpen
  \bibfield  {author} {\bibinfo {author} {\bibfnamefont {I.~L.}\ \bibnamefont
  {Medintz}}\ and\ \bibinfo {author} {\bibfnamefont {N.}~\bibnamefont
  {Hildebrandt}},\ }\href@noop {} {\emph {\bibinfo {title} {FRET - F\"orster
  Resonance Energy Transfer: From Theory to Applications}}}\ (\bibinfo
  {publisher} {John Wiley \& Sons},\ \bibinfo {year} {2003})\BibitemShut
  {NoStop}%
\bibitem [{\citenamefont {Ha}(2001)}]{Ha2001}%
  \BibitemOpen
  \bibfield  {author} {\bibinfo {author} {\bibfnamefont {T.}~\bibnamefont
  {Ha}},\ }\href {\doibase http://dx.doi.org/10.1006/meth.2001.1217} {\bibfield
   {journal} {\bibinfo  {journal} {Methods}\ }\textbf {\bibinfo {volume}
  {25}},\ \bibinfo {pages} {78} (\bibinfo {year} {2001})}\BibitemShut {NoStop}%
\bibitem [{\citenamefont {Carminati}\ \emph {et~al.}(2015)\citenamefont
  {Carminati}, \citenamefont {Caz{\'e}}, \citenamefont {Cao}, \citenamefont
  {Peragut}, \citenamefont {Krachmalnicoff}, \citenamefont {Pierrat},\ and\
  \citenamefont {Wilde}}]{Carminati2015}%
  \BibitemOpen
  \bibfield  {author} {\bibinfo {author} {\bibfnamefont {R.}~\bibnamefont
  {Carminati}}, \bibinfo {author} {\bibfnamefont {A.}~\bibnamefont {Caz{\'e}}},
  \bibinfo {author} {\bibfnamefont {D.}~\bibnamefont {Cao}}, \bibinfo {author}
  {\bibfnamefont {F.}~\bibnamefont {Peragut}}, \bibinfo {author} {\bibfnamefont
  {V.}~\bibnamefont {Krachmalnicoff}}, \bibinfo {author} {\bibfnamefont
  {R.}~\bibnamefont {Pierrat}}, \ and\ \bibinfo {author} {\bibfnamefont
  {Y.~D.}\ \bibnamefont {Wilde}},\ }\href {\doibase
  http://dx.doi.org/10.1016/j.surfrep.2014.11.001} {\bibfield  {journal}
  {\bibinfo  {journal} {Surface Science Reports}\ }\textbf {\bibinfo {volume}
  {70}},\ \bibinfo {pages} {1 } (\bibinfo {year} {2015})}\BibitemShut {NoStop}%
\bibitem [{\citenamefont {Koenderink}\ \emph {et~al.}(2015)\citenamefont
  {Koenderink}, \citenamefont {Al\'u},\ and\ \citenamefont
  {Polman}}]{Koenderink2015}%
  \BibitemOpen
  \bibfield  {author} {\bibinfo {author} {\bibfnamefont {A.~F.}\ \bibnamefont
  {Koenderink}}, \bibinfo {author} {\bibfnamefont {A.}~\bibnamefont {Al\'u}}, \
  and\ \bibinfo {author} {\bibfnamefont {A.}~\bibnamefont {Polman}},\ }\href
  {\doibase 10.1126/science.1261243} {\bibfield  {journal} {\bibinfo  {journal}
  {Science}\ }\textbf {\bibinfo {volume} {348}},\ \bibinfo {pages} {516}
  (\bibinfo {year} {2015})}\BibitemShut {NoStop}%
\bibitem [{\citenamefont {Curto}\ \emph {et~al.}(2010)\citenamefont {Curto},
  \citenamefont {Volpe}, \citenamefont {Taminiau}, \citenamefont {Kreuzer},
  \citenamefont {Quidant},\ and\ \citenamefont {van Hulst}}]{Curto2010}%
  \BibitemOpen
  \bibfield  {author} {\bibinfo {author} {\bibfnamefont {A.~G.}\ \bibnamefont
  {Curto}}, \bibinfo {author} {\bibfnamefont {G.}~\bibnamefont {Volpe}},
  \bibinfo {author} {\bibfnamefont {T.~H.}\ \bibnamefont {Taminiau}}, \bibinfo
  {author} {\bibfnamefont {M.~P.}\ \bibnamefont {Kreuzer}}, \bibinfo {author}
  {\bibfnamefont {R.}~\bibnamefont {Quidant}}, \ and\ \bibinfo {author}
  {\bibfnamefont {N.~F.}\ \bibnamefont {van Hulst}},\ }\href {\doibase
  10.1126/science.1191922} {\bibfield  {journal} {\bibinfo  {journal}
  {Science}\ }\textbf {\bibinfo {volume} {329}},\ \bibinfo {pages} {930}
  (\bibinfo {year} {2010})}\BibitemShut {NoStop}%
\bibitem [{\citenamefont {Tame}\ \emph {et~al.}(2013)\citenamefont {Tame},
  \citenamefont {McEnery}, \citenamefont {Ozdemir}, \citenamefont {Lee},
  \citenamefont {Maier},\ and\ \citenamefont {Kim}}]{Tame2013}%
  \BibitemOpen
  \bibfield  {author} {\bibinfo {author} {\bibfnamefont {M.~S.}\ \bibnamefont
  {Tame}}, \bibinfo {author} {\bibfnamefont {K.~R.}\ \bibnamefont {McEnery}},
  \bibinfo {author} {\bibfnamefont {S.~K.}\ \bibnamefont {Ozdemir}}, \bibinfo
  {author} {\bibfnamefont {J.}~\bibnamefont {Lee}}, \bibinfo {author}
  {\bibfnamefont {S.~A.}\ \bibnamefont {Maier}}, \ and\ \bibinfo {author}
  {\bibfnamefont {M.~S.}\ \bibnamefont {Kim}},\ }\href
  {http://dx.doi.org/10.1038/nphys2615} {\bibfield  {journal} {\bibinfo
  {journal} {Nat Phys}\ }\textbf {\bibinfo {volume} {9}},\ \bibinfo {pages}
  {329} (\bibinfo {year} {2013})}\BibitemShut {NoStop}%
\bibitem [{\citenamefont {Kolesov}\ \emph {et~al.}(2009)\citenamefont
  {Kolesov}, \citenamefont {Grotz}, \citenamefont {Balasubramanian},
  \citenamefont {Stohr}, \citenamefont {Nicolet}, \citenamefont {Hemmer},
  \citenamefont {Jelezko},\ and\ \citenamefont {Wrachtrup}}]{Wrachtrup2009}%
  \BibitemOpen
  \bibfield  {author} {\bibinfo {author} {\bibfnamefont {R.}~\bibnamefont
  {Kolesov}}, \bibinfo {author} {\bibfnamefont {B.}~\bibnamefont {Grotz}},
  \bibinfo {author} {\bibfnamefont {G.}~\bibnamefont {Balasubramanian}},
  \bibinfo {author} {\bibfnamefont {R.~J.}\ \bibnamefont {Stohr}}, \bibinfo
  {author} {\bibfnamefont {A.~A.~L.}\ \bibnamefont {Nicolet}}, \bibinfo
  {author} {\bibfnamefont {P.~R.}\ \bibnamefont {Hemmer}}, \bibinfo {author}
  {\bibfnamefont {F.}~\bibnamefont {Jelezko}}, \ and\ \bibinfo {author}
  {\bibfnamefont {J.}~\bibnamefont {Wrachtrup}},\ }\href
  {http://dx.doi.org/10.1038/nphys1278} {\bibfield  {journal} {\bibinfo
  {journal} {Nat. Phys.}\ }\textbf {\bibinfo {volume} {5}},\ \bibinfo {pages}
  {470} (\bibinfo {year} {2009})}\BibitemShut {NoStop}%
\bibitem [{\citenamefont {Fakonas}\ \emph {et~al.}(2014)\citenamefont
  {Fakonas}, \citenamefont {Lee}, \citenamefont {Kelaita},\ and\ \citenamefont
  {Atwater}}]{Fakonas2014}%
  \BibitemOpen
  \bibfield  {author} {\bibinfo {author} {\bibfnamefont {J.~S.}\ \bibnamefont
  {Fakonas}}, \bibinfo {author} {\bibfnamefont {H.}~\bibnamefont {Lee}},
  \bibinfo {author} {\bibfnamefont {Y.~A.}\ \bibnamefont {Kelaita}}, \ and\
  \bibinfo {author} {\bibfnamefont {H.~A.}\ \bibnamefont {Atwater}},\ }\href
  {http://dx.doi.org/10.1038/nphoton.2014.40} {\bibfield  {journal} {\bibinfo
  {journal} {Nat. Photon.}\ }\textbf {\bibinfo {volume} {8}},\ \bibinfo {pages}
  {317} (\bibinfo {year} {2014})}\BibitemShut {NoStop}%
\bibitem [{\citenamefont {Akselrod}\ \emph {et~al.}(2014)\citenamefont
  {Akselrod}, \citenamefont {Argyropoulos}, \citenamefont {Hoang},
  \citenamefont {Cirac{\`\i}}, \citenamefont {Fang}, \citenamefont {Huang},
  \citenamefont {Smith},\ and\ \citenamefont {Mikkelsen}}]{Akselrod2014}%
  \BibitemOpen
  \bibfield  {author} {\bibinfo {author} {\bibfnamefont {G.~M.}\ \bibnamefont
  {Akselrod}}, \bibinfo {author} {\bibfnamefont {C.}~\bibnamefont
  {Argyropoulos}}, \bibinfo {author} {\bibfnamefont {T.~B.}\ \bibnamefont
  {Hoang}}, \bibinfo {author} {\bibfnamefont {C.}~\bibnamefont {Cirac{\`\i}}},
  \bibinfo {author} {\bibfnamefont {C.}~\bibnamefont {Fang}}, \bibinfo {author}
  {\bibfnamefont {J.}~\bibnamefont {Huang}}, \bibinfo {author} {\bibfnamefont
  {D.~R.}\ \bibnamefont {Smith}}, \ and\ \bibinfo {author} {\bibfnamefont
  {M.~H.}\ \bibnamefont {Mikkelsen}},\ }\href@noop {} {\bibfield  {journal}
  {\bibinfo  {journal} {Nat. Photon.}\ }\textbf {\bibinfo {volume} {8}},\
  \bibinfo {pages} {835} (\bibinfo {year} {2014})}\BibitemShut {NoStop}%
\bibitem [{\citenamefont {Belacel}\ \emph {et~al.}(2013)\citenamefont
  {Belacel}, \citenamefont {Habert}, \citenamefont {Bigourdan}, \citenamefont
  {Marquier}, \citenamefont {Hugonin}, \citenamefont {Michaelis~de
  Vasconcellos}, \citenamefont {Lafosse}, \citenamefont {Coolen}, \citenamefont
  {Schwob}, \citenamefont {Javaux}, \citenamefont {Dubertret}, \citenamefont
  {Greffet}, \citenamefont {Senellart},\ and\ \citenamefont
  {Maitre}}]{Belacel2013}%
  \BibitemOpen
  \bibfield  {author} {\bibinfo {author} {\bibfnamefont {C.}~\bibnamefont
  {Belacel}}, \bibinfo {author} {\bibfnamefont {B.}~\bibnamefont {Habert}},
  \bibinfo {author} {\bibfnamefont {F.}~\bibnamefont {Bigourdan}}, \bibinfo
  {author} {\bibfnamefont {F.}~\bibnamefont {Marquier}}, \bibinfo {author}
  {\bibfnamefont {J.-P.}\ \bibnamefont {Hugonin}}, \bibinfo {author}
  {\bibfnamefont {S.}~\bibnamefont {Michaelis~de Vasconcellos}}, \bibinfo
  {author} {\bibfnamefont {X.}~\bibnamefont {Lafosse}}, \bibinfo {author}
  {\bibfnamefont {L.}~\bibnamefont {Coolen}}, \bibinfo {author} {\bibfnamefont
  {C.}~\bibnamefont {Schwob}}, \bibinfo {author} {\bibfnamefont
  {C.}~\bibnamefont {Javaux}}, \bibinfo {author} {\bibfnamefont
  {B.}~\bibnamefont {Dubertret}}, \bibinfo {author} {\bibfnamefont {J.-J.}\
  \bibnamefont {Greffet}}, \bibinfo {author} {\bibfnamefont {P.}~\bibnamefont
  {Senellart}}, \ and\ \bibinfo {author} {\bibfnamefont {A.}~\bibnamefont
  {Maitre}},\ }\href {\doibase 10.1021/nl3046602} {\bibfield  {journal}
  {\bibinfo  {journal} {Nano Letters}\ }\textbf {\bibinfo {volume} {13}},\
  \bibinfo {pages} {1516} (\bibinfo {year} {2013})}\BibitemShut {NoStop}%
\bibitem [{\citenamefont {Leitner}\ and\ \citenamefont
  {Reinish}(1988)}]{Leitner1988}%
  \BibitemOpen
  \bibfield  {author} {\bibinfo {author} {\bibfnamefont {A.}~\bibnamefont
  {Leitner}}\ and\ \bibinfo {author} {\bibfnamefont {H.}~\bibnamefont
  {Reinish}},\ }\href@noop {} {\bibfield  {journal} {\bibinfo  {journal} {Chem.
  Phys. Lett.}\ }\textbf {\bibinfo {volume} {146}},\ \bibinfo {pages} {320}
  (\bibinfo {year} {1988})}\BibitemShut {NoStop}%
\bibitem [{\citenamefont {de~Dood}\ \emph {et~al.}(2005)\citenamefont
  {de~Dood}, \citenamefont {Knoester}, \citenamefont {Tip},\ and\ \citenamefont
  {Polman}}]{Dood2005}%
  \BibitemOpen
  \bibfield  {author} {\bibinfo {author} {\bibfnamefont {M.}~\bibnamefont
  {de~Dood}}, \bibinfo {author} {\bibfnamefont {J.}~\bibnamefont {Knoester}},
  \bibinfo {author} {\bibfnamefont {A.}~\bibnamefont {Tip}}, \ and\ \bibinfo
  {author} {\bibfnamefont {A.}~\bibnamefont {Polman}},\ }\href@noop {}
  {\bibfield  {journal} {\bibinfo  {journal} {Phys. Rev. B}\ }\textbf {\bibinfo
  {volume} {71}},\ \bibinfo {pages} {115102} (\bibinfo {year}
  {2005})}\BibitemShut {NoStop}%
\bibitem [{\citenamefont {Nakamura}\ \emph {et~al.}(2006)\citenamefont
  {Nakamura}, \citenamefont {Fujii}, \citenamefont {Miura}, \citenamefont
  {Inui},\ and\ \citenamefont {Hayashi}}]{Nakamura2006}%
  \BibitemOpen
  \bibfield  {author} {\bibinfo {author} {\bibfnamefont {T.}~\bibnamefont
  {Nakamura}}, \bibinfo {author} {\bibfnamefont {M.}~\bibnamefont {Fujii}},
  \bibinfo {author} {\bibfnamefont {S.}~\bibnamefont {Miura}}, \bibinfo
  {author} {\bibfnamefont {M.}~\bibnamefont {Inui}}, \ and\ \bibinfo {author}
  {\bibfnamefont {S.}~\bibnamefont {Hayashi}},\ }\href@noop {} {\bibfield
  {journal} {\bibinfo  {journal} {Phys. Rev. B}\ }\textbf {\bibinfo {volume}
  {74}},\ \bibinfo {pages} {045302} (\bibinfo {year} {2006})}\BibitemShut
  {NoStop}%
\bibitem [{\citenamefont {Zhang}\ \emph {et~al.}(2007)\citenamefont {Zhang},
  \citenamefont {Fu},\ and\ \citenamefont {Lakowicz}}]{Zhang2007}%
  \BibitemOpen
  \bibfield  {author} {\bibinfo {author} {\bibfnamefont {J.}~\bibnamefont
  {Zhang}}, \bibinfo {author} {\bibfnamefont {Y.}~\bibnamefont {Fu}}, \ and\
  \bibinfo {author} {\bibfnamefont {J.~R.}\ \bibnamefont {Lakowicz}},\
  }\href@noop {} {\bibfield  {journal} {\bibinfo  {journal} {J. of Phys. Chem.
  C}\ }\textbf {\bibinfo {volume} {111}},\ \bibinfo {pages} {50} (\bibinfo
  {year} {2007})}\BibitemShut {NoStop}%
\bibitem [{\citenamefont {Komarala}\ \emph {et~al.}(2008)\citenamefont
  {Komarala}, \citenamefont {Bradley}, \citenamefont {Rakovich}, \citenamefont
  {Byrne}, \citenamefont {Gun'ko},\ and\ \citenamefont
  {Rogach}}]{Komarala2008}%
  \BibitemOpen
  \bibfield  {author} {\bibinfo {author} {\bibfnamefont {V.~K.}\ \bibnamefont
  {Komarala}}, \bibinfo {author} {\bibfnamefont {A.~L.}\ \bibnamefont
  {Bradley}}, \bibinfo {author} {\bibfnamefont {Y.~P.}\ \bibnamefont
  {Rakovich}}, \bibinfo {author} {\bibfnamefont {S.~J.}\ \bibnamefont {Byrne}},
  \bibinfo {author} {\bibfnamefont {Y.~K.}\ \bibnamefont {Gun'ko}}, \ and\
  \bibinfo {author} {\bibfnamefont {A.~L.}\ \bibnamefont {Rogach}},\
  }\href@noop {} {\bibfield  {journal} {\bibinfo  {journal} {Applied Physics
  Letters}\ }\textbf {\bibinfo {volume} {93}},\ \bibinfo {pages} {123102}
  (\bibinfo {year} {2008})}\BibitemShut {NoStop}%
\bibitem [{\citenamefont {Reil}\ \emph {et~al.}(2008)\citenamefont {Reil},
  \citenamefont {Hohenester}, \citenamefont {Krenn},\ and\ \citenamefont
  {Leitner}}]{Reil2008}%
  \BibitemOpen
  \bibfield  {author} {\bibinfo {author} {\bibfnamefont {F.}~\bibnamefont
  {Reil}}, \bibinfo {author} {\bibfnamefont {U.}~\bibnamefont {Hohenester}},
  \bibinfo {author} {\bibfnamefont {J.~R.}\ \bibnamefont {Krenn}}, \ and\
  \bibinfo {author} {\bibfnamefont {A.}~\bibnamefont {Leitner}},\ }\href
  {\doibase 10.1021/nl801480m} {\bibfield  {journal} {\bibinfo  {journal} {Nano
  Letters}\ }\textbf {\bibinfo {volume} {8}},\ \bibinfo {pages} {4128}
  (\bibinfo {year} {2008})}\BibitemShut {NoStop}%
\bibitem [{\citenamefont {Faessler}\ \emph {et~al.}(2011)\citenamefont
  {Faessler}, \citenamefont {Hrelescu}, \citenamefont {Lutich}, \citenamefont
  {Osinkina}, \citenamefont {Mayilo}, \citenamefont {J{\"a}ckel},\ and\
  \citenamefont {Feldmann}}]{Faessler2011}%
  \BibitemOpen
  \bibfield  {author} {\bibinfo {author} {\bibfnamefont {V.}~\bibnamefont
  {Faessler}}, \bibinfo {author} {\bibfnamefont {C.}~\bibnamefont {Hrelescu}},
  \bibinfo {author} {\bibfnamefont {A.~A.}\ \bibnamefont {Lutich}}, \bibinfo
  {author} {\bibfnamefont {L.}~\bibnamefont {Osinkina}}, \bibinfo {author}
  {\bibfnamefont {S.}~\bibnamefont {Mayilo}}, \bibinfo {author} {\bibfnamefont
  {F.}~\bibnamefont {J{\"a}ckel}}, \ and\ \bibinfo {author} {\bibfnamefont
  {J.}~\bibnamefont {Feldmann}},\ }\href@noop {} {\bibfield  {journal}
  {\bibinfo  {journal} {Chem. Phys. Lett.}\ }\textbf {\bibinfo {volume}
  {508}},\ \bibinfo {pages} {67} (\bibinfo {year} {2011})}\BibitemShut
  {NoStop}%
\bibitem [{\citenamefont {Blum}\ \emph {et~al.}(2012)\citenamefont {Blum},
  \citenamefont {Zijlstra}, \citenamefont {Lagendijk}, \citenamefont {Wubs},
  \citenamefont {Mosk}, \citenamefont {Subramaniam},\ and\ \citenamefont
  {Vos}}]{Blum2012}%
  \BibitemOpen
  \bibfield  {author} {\bibinfo {author} {\bibfnamefont {C.}~\bibnamefont
  {Blum}}, \bibinfo {author} {\bibfnamefont {N.}~\bibnamefont {Zijlstra}},
  \bibinfo {author} {\bibfnamefont {A.}~\bibnamefont {Lagendijk}}, \bibinfo
  {author} {\bibfnamefont {M.}~\bibnamefont {Wubs}}, \bibinfo {author}
  {\bibfnamefont {A.~P.}\ \bibnamefont {Mosk}}, \bibinfo {author}
  {\bibfnamefont {V.}~\bibnamefont {Subramaniam}}, \ and\ \bibinfo {author}
  {\bibfnamefont {W.~L.}\ \bibnamefont {Vos}},\ }\href@noop {} {\bibfield
  {journal} {\bibinfo  {journal} {Phys. Rev. Lett.}\ }\textbf {\bibinfo
  {volume} {109}},\ \bibinfo {pages} {203601} (\bibinfo {year}
  {2012})}\BibitemShut {NoStop}%
\bibitem [{\citenamefont {Ghenuche}\ \emph {et~al.}(2014)\citenamefont
  {Ghenuche}, \citenamefont {de~Torres}, \citenamefont {Moparthi},
  \citenamefont {Grigoriev},\ and\ \citenamefont {Wenger}}]{Ghenuche2014}%
  \BibitemOpen
  \bibfield  {author} {\bibinfo {author} {\bibfnamefont {P.}~\bibnamefont
  {Ghenuche}}, \bibinfo {author} {\bibfnamefont {J.}~\bibnamefont {de~Torres}},
  \bibinfo {author} {\bibfnamefont {S.~B.}\ \bibnamefont {Moparthi}}, \bibinfo
  {author} {\bibfnamefont {V.}~\bibnamefont {Grigoriev}}, \ and\ \bibinfo
  {author} {\bibfnamefont {J.}~\bibnamefont {Wenger}},\ }\href {\doibase
  10.1021/nl5018145} {\bibfield  {journal} {\bibinfo  {journal} {Nano Letters}\
  }\textbf {\bibinfo {volume} {14}},\ \bibinfo {pages} {4707} (\bibinfo {year}
  {2014})}\BibitemShut {NoStop}%
\bibitem [{\citenamefont {de~Torres}\ \emph {et~al.}(2015)\citenamefont
  {de~Torres}, \citenamefont {Ghenuche}, \citenamefont {Moparthi},
  \citenamefont {Grigoriev},\ and\ \citenamefont {Wenger}}]{Torres2015}%
  \BibitemOpen
  \bibfield  {author} {\bibinfo {author} {\bibfnamefont {J.}~\bibnamefont
  {de~Torres}}, \bibinfo {author} {\bibfnamefont {P.}~\bibnamefont {Ghenuche}},
  \bibinfo {author} {\bibfnamefont {S.~B.}\ \bibnamefont {Moparthi}}, \bibinfo
  {author} {\bibfnamefont {V.}~\bibnamefont {Grigoriev}}, \ and\ \bibinfo
  {author} {\bibfnamefont {J.}~\bibnamefont {Wenger}},\ }\href {\doibase
  10.1002/cphc.201402651} {\bibfield  {journal} {\bibinfo  {journal}
  {ChemPhysChem}\ }\textbf {\bibinfo {volume} {16}},\ \bibinfo {pages} {782}
  (\bibinfo {year} {2015})}\BibitemShut {NoStop}%
\bibitem [{\citenamefont {Raki\'{c}}\ \emph {et~al.}(1998)\citenamefont
  {Raki\'{c}}, \citenamefont {Djuri\v{s}i\'{c}}, \citenamefont {Elazar},\ and\
  \citenamefont {Majewski}}]{Rakic1998}%
  \BibitemOpen
  \bibfield  {author} {\bibinfo {author} {\bibfnamefont {A.~D.}\ \bibnamefont
  {Raki\'{c}}}, \bibinfo {author} {\bibfnamefont {A.~B.}\ \bibnamefont
  {Djuri\v{s}i\'{c}}}, \bibinfo {author} {\bibfnamefont {J.~M.}\ \bibnamefont
  {Elazar}}, \ and\ \bibinfo {author} {\bibfnamefont {M.~L.}\ \bibnamefont
  {Majewski}},\ }\href {\doibase 10.1364/AO.37.005271} {\bibfield  {journal}
  {\bibinfo  {journal} {Appl. Opt.}\ }\textbf {\bibinfo {volume} {37}},\
  \bibinfo {pages} {5271} (\bibinfo {year} {1998})}\BibitemShut {NoStop}%
\bibitem{Supplemental_Material}See Supplemental Material at [URL will be inserted by publisher], which includes Refs. [4,22,28,29].    
\bibitem [{\citenamefont {Andrew}\ and\ \citenamefont
  {Barnes}(2004)}]{Andrew2004}%
  \BibitemOpen
  \bibfield  {author} {\bibinfo {author} {\bibfnamefont {P.}~\bibnamefont
  {Andrew}}\ and\ \bibinfo {author} {\bibfnamefont {W.~L.}\ \bibnamefont
  {Barnes}},\ }\href@noop {} {\bibfield  {journal} {\bibinfo  {journal}
  {Science}\ }\textbf {\bibinfo {volume} {306}},\ \bibinfo {pages} {1002}
  (\bibinfo {year} {2004})}\BibitemShut {NoStop}%
\bibitem{Marocico2011}
C.A. Marocico, J. Knoester, Phys. Rev. A {\bf 84}, 053824 (2011).  
\bibitem [{\citenamefont {Krachmalnicoff}\ \emph {et~al.}(2013)\citenamefont
  {Krachmalnicoff}, \citenamefont {Cao}, \citenamefont {Caz\'{e}},
  \citenamefont {Castani\'{e}}, \citenamefont {Pierrat}, \citenamefont
  {Bardou}, \citenamefont {Collin}, \citenamefont {Carminati},\ and\
  \citenamefont {Wilde}}]{Krachmalnicoff2013}%
  \BibitemOpen
  \bibfield  {author} {\bibinfo {author} {\bibfnamefont {V.}~\bibnamefont
  {Krachmalnicoff}}, \bibinfo {author} {\bibfnamefont {D.}~\bibnamefont {Cao}},
  \bibinfo {author} {\bibfnamefont {A.}~\bibnamefont {Caz\'{e}}}, \bibinfo
  {author} {\bibfnamefont {E.}~\bibnamefont {Castani\'{e}}}, \bibinfo {author}
  {\bibfnamefont {R.}~\bibnamefont {Pierrat}}, \bibinfo {author} {\bibfnamefont
  {N.}~\bibnamefont {Bardou}}, \bibinfo {author} {\bibfnamefont
  {S.}~\bibnamefont {Collin}}, \bibinfo {author} {\bibfnamefont
  {R.}~\bibnamefont {Carminati}}, \ and\ \bibinfo {author} {\bibfnamefont
  {Y.~D.}\ \bibnamefont {Wilde}},\ }\href {\doibase 10.1364/OE.21.011536}
  {\bibfield  {journal} {\bibinfo  {journal} {Opt. Express}\ }\textbf {\bibinfo
  {volume} {21}},\ \bibinfo {pages} {11536} (\bibinfo {year}
  {2013})}\BibitemShut {NoStop}%
\bibitem [{\citenamefont {Cao}\ \emph {et~al.}(2015)\citenamefont {Cao},
  \citenamefont {Caz\'e¬é}, \citenamefont {Calabrese}, \citenamefont
  {Pierrat}, \citenamefont {Bardou}, \citenamefont {Collin}, \citenamefont
  {Carminati}, \citenamefont {Krachmalnicoff},\ and\ \citenamefont
  {De~Wilde}}]{Cao2015}%
  \BibitemOpen
  \bibfield  {author} {\bibinfo {author} {\bibfnamefont {D.}~\bibnamefont
  {Cao}}, \bibinfo {author} {\bibfnamefont {A.}~\bibnamefont {Caz\'e¬é}},
  \bibinfo {author} {\bibfnamefont {M.}~\bibnamefont {Calabrese}}, \bibinfo
  {author} {\bibfnamefont {R.}~\bibnamefont {Pierrat}}, \bibinfo {author}
  {\bibfnamefont {N.}~\bibnamefont {Bardou}}, \bibinfo {author} {\bibfnamefont
  {S.}~\bibnamefont {Collin}}, \bibinfo {author} {\bibfnamefont
  {R.}~\bibnamefont {Carminati}}, \bibinfo {author} {\bibfnamefont
  {V.}~\bibnamefont {Krachmalnicoff}}, \ and\ \bibinfo {author} {\bibfnamefont
  {Y.}~\bibnamefont {De~Wilde}},\ }\href {\doibase 10.1021/ph500431g}
  {\bibfield  {journal} {\bibinfo  {journal} {ACS Photonics}\ }\textbf
  {\bibinfo {volume} {2}},\ \bibinfo {pages} {189} (\bibinfo {year}
  {2015})}\BibitemShut {NoStop}%  
\bibitem{novotny_principles_2006}
L. Novotny and B. Hecht, Principles of Nano-Optics (Cambridge University, 2006).
\bibitem{dung_intermolecular_2002}
H.~T.~Dung, L.~Kn\"oll, and D.-G. Welsch, Phys. Rev. A, {\bf 65},043813 (2002).
\end{thebibliography}
\end{document}